\documentstyle[12pt]{article}
\newcommand{\bce}{\begin{center}}
\newcommand{\ece}{\end{center}}
\newcommand{\beq}{\begin{equation}}
\newcommand{\eeq}{\end{equation}}
\newcommand{\bea}{\vspace{0.25cm}\begin{eqnarray}}
\newcommand{\eea}{\end{eqnarray}}

\newcommand{\ba}{\begin{array}}
\newcommand{\ea}{\end{array}}

\newcommand{\doublespace}{
    \renewcommand{\baselinestretch}{1.6}\large\normalsize}

\def\lsim{\mathrel{\rlap{\lower4pt\hbox{\hskip1pt$\sim$}}
    \raise1pt\hbox{$<$}}}     
\def\gsim{\mathrel{\rlap{\lower4pt\hbox{\hskip1pt$\sim$}}
    \raise1pt\hbox{$>$}}}     

\def\lsim{\mathrel{\rlap{\lower4pt\hbox{\hskip1pt$\sim$}}
    \raise1pt\hbox{$<$}}}         
\def\gsim{\mathrel{\rlap{\lower4pt\hbox{\hskip1pt$\sim$}}
    \raise1pt\hbox{$>$}}}         

\def\beq{\begin{equation}}
\def\endeq{\end{equation}}
\def\arr{\begin{eqnarray}}
\def\endarr{\end{eqnarray}}

\makeindex
\begin{document}
\phantom{.}\hspace{12.0cm}{July 23, 1993}
\vspace{2cm}
\begin{center}
{\bf \huge Theoretical Interpretation of the NE18 Experiment
on Nuclear Transparency in  $A(e,e'p)$ Scattering \\}
\vspace{1cm}
{\bf N.N.Nikolaev$^{1,2)}$, A.Szczurek$^{1,3)}$,
J.Speth$^{1)}$, J.Wambach$^{1,4)}$, \\
B.G.Zakharov$^{2)}$ and V.R.Zoller$^{5)}$ } \medskip\\
{\small \sl
$^{1)}$IKP(Theorie), Forschungszentrum  J\"ulich GmbH.,\\
D-52425 J\"ulich, Germany \\
$^{2)}$L.D.Landau Institute for Theoretical Physics, \\
GSP-1, 117940, ul.Kosygina 2, V-334 Moscow, Russia\\
$^{3)}$ Institute of Nuclear Physics, PL-31-342 Krakow, Poland\\
$^{4)}$Department of Physics, University of Illinois at
Urbana-Champaign, \\
Urbana, IL 61801, USA\\
$^{5)}$Institute for Theoretical and Experimental Physics, \\
ul.B.Cheremushkinskaya 25, 117259 Moscow, Russia
\vspace{1cm}\\}
{\bf \LARGE A b s t r a c t \bigskip\\}
\end{center}

The spectral function, measured
in $A(e,e'p)$ reactions, is distorted by
the final-state interaction of the struck proton with the
residual nucleus.
This causes a broadening of the observed transverse-momentum
distribution which is large even in the $d(e,e'p)$ reaction.
We discuss the effects of this $p_{\perp}$-broadening on
the nuclear transparency measured in the recent NE18 experiment.
Within conventional Glauber theory we can describe  the
measurements. Transparency effects are thus small in agreement with
our earlier predictions.

\newpage
\doublespace

The recently completed
SLAC NE18 experiment [1] on $A(e,e'p)$ scattering
concludes that, up to $Q^{2} \lsim 7$ (GeV/c)$^{2}$,
there is no conclusive evidence for color transparency (CT).
A slow onset of CT was predicted by us [2,3] in an
approach which is based on CT sum rules and the quark-hadron
duality. It was shown that CT and/or weak final-state
interaction (FSI) in $A(e,e'p)$ reactions arises from
delicate cancellations of contributions
from elastic ($|i\rangle=|p\rangle$) and inelastic
($|i\rangle \neq |p\rangle$ ) intermediate states $|i\rangle$
propagating
inside the nucleus [2,3].
At $Q^{2}\lsim 7\,{\rm (GeV/c)}^{2}$ the contributions from inelastic
rescatterings is still small, however, and hence the very
small signal in the NE18 experiment (for a general review on CT see
[5-7]). Another very
interesting aspect of the experiment is the evidence
for substantial FSI effects already in the deuteron.

A quantitative description of the NE18 data [1] is an
important check for our understanding of the FSI at large
$Q^{2}$ and the purpose of this communication is to provide such
a description. To do so, one
should realize that, because of the finite
spectrometer acceptance in the
missing energy $E_{m}$ and missing momentum
$\vec{p}_{m}=(p_{m,z},\vec{p}_{\perp})$
(the $z$-axis is chosen along
the $\vec{q}$ direction), the sum-rule
technique of refs.~[2,3] is not
applicable. It has been argued in [1] that
one should compare the (partially integrated)
measured spectral function
$S(E_{m},\vec{p}_{m})$ to the calculated $S_{PWIA}(E_{m},\vec{p}_{m})$
in the plane-wave impulse approximation (PWIA). We will show,
however, that the spectral function does not factorize as
$S_{PWIA}(E_{m},\vec{p}_{m})$ times
a $(E_{m},\vec{p}_{m})$-independent
attenuation factor. This is due to the fact the FSI of
the struck proton
significantly distorts the $p_{\perp}$-distribution [8].
Evidently, elastic rescatterings
deflect the struck proton and only the inelastic $pN$ cross
section $\sigma_{in}(pN)$ contributes to the attenuation of the
$p_{\perp}$-integrated cross section [2,8]. In parallel kinematics
($\vec{p}_{\perp}=0$), on the other hand, the attenuation
is controlled by $\sigma_{tot}(pN)$ [8]. Thus,
because of the finite $p_{\perp}$-acceptance in the NE18 experiment,
an accurate evaluation of the FSI effects is necessary
for the quantitative
interpretation of the data.

We begin with the analysis of the $d(e,e'p)$ scattering. At
$Q^{2}\gsim 1\,{\rm (GeV/c)}^{2}$ the kinetic energy of
the struck proton is large, $T_{kin}\approx Q^{2}/2m_{p}$, and
we can use the Glauber approximation for the FSI of the
struck proton with the spectator neutron [8]. Then the momentum
distribution, $f_d(\vec p_{m})$, of the observed protons can be
written as
\beq
f_d(p_{m,z},\vec{p}_{\perp})=
\left|\phi_{d}(p_{m,z},\vec{p}_{\perp})-
{\sigma_{tot}(pn) \over 16\pi^{2}}\int d^{2}k\,
\phi_{d}(p_{m,z},\vec{p}_{\perp}-\vec{k})
\exp\left(-{1\over 2}B\vec{k}^{2}\right)
\right|^{2} \, ,
\label{eq:1}
\endeq
where $\phi_{d}(\vec{k})$ is the momentum-space wave function
of the deuteron, and $B$ denotes the diffraction slope for elastic $pn$
scattering. The measured transparency factor
\beq
T_{d}(p_{m,z}=0)=
{\int^{p_\perp^{max}}d^{2}p_{\perp}f_d(\vec{p}_{\perp})
/
\int^{p_\perp^{max}}d^{2}\vec{p}_{\perp}|\phi_d(\vec{p}_{\perp})|^{2} }
\label{eq:2}
\eeq
explicitly depends on the values $p_{\perp} \leq p_\perp^{max}$ accepted
in the spectrometer. In NE18, $p_{\perp}^{max}=170$ MeV/c while the
$p_{m,z}$-acceptance is very small [1]. Since
$B \ll R_{d}^{2}$, where $R_{d}$ denotes the radius of the deuteron,
the rescattering term in (\ref{eq:1}) is almost constant over this
range of $p_{\perp}$. On the other hand,
$\left(R_{d}p_{\perp}^{max}\right)^{2} \gg 1$.
This leads to a simple estimate of the attenuation effect
\beq
1-T_{d} \sim {\sigma_{tot}(pn) \over 2\pi R_{d}^{2}} \sim 0.07\,\, ,
\label{eq:3}
\endeq
which is about twice as large as the Glauber's shadowing effect in
$\sigma_{tot}(pd)$ [9,10]. Using a realistic Bonn wave function for
the deuteron [11], the detailed predictions at $p_{m,z}=0$ are shown in
Fig.~1.

An  extension to heavier targets is straightforward.
With the broad
$E_{m}$-acceptance of the NE18 experiment closure becomes
applicable and, in the absence of FSI, one would obtain the
single-particle momentum distribution,
$d\sigma_{PWIA}\propto n_{F}(\vec{p})$ with
\beq
n_{F}(\vec{p})= {1\over Z}\int dE_{m}S_{PWIA}(E_{m},\vec{p})=
{1\over A(2\pi)^{3}}
\int d\vec{r}_{1}\,d\vec{r}_{1}\,'\,
\rho_{1}(\vec{r}_{1},\vec{r}_{1}\,')
\exp[i\vec{k}'(\vec{r}_{1}-\vec{r}_{1}\,')]\,\, ,
\eeq
where  $\rho_{1}(\vec{r},\vec{r}\,')$
denotes the one-body
density matrix while $Z$ and $A$ are the charge and
mass number of the nucleus,
respectively. With allowance for FSI [6,8,12],
on the other hand,
\arr
d\sigma_{A} \propto
{1\over Z}\int dE_{m}S(E_{m},\vec{p}_{m})=
\int d\vec{r}\,'d\vec{r} \,
\rho_{1}(\vec{r},\vec{r}\,')
\exp[i\vec{p}_{m}(\vec{r}\,'-\vec{r})] ~~~~~~~\nonumber\\
\cdot \exp\left[
-{1\over 2}(1-i\alpha_{pN})\sigma_{tot}(pN)
t(\vec{b},z)
-{1\over 2}(1+i\alpha_{pN})\sigma_{tot}(pN)
t(\vec{b}',z') \right] \nonumber\\
\cdot\exp\left[ t(\vec{b},max(z,z')) \xi(\vec{\Delta}) \right]\,. ~~~~~~~
{}~~~~~~
\label{eq:4}
\endarr
where $\vec{r}=(\vec{b},z),\,\vec{r}\,'=(\vec{b}',z')$,
$\vec{\Delta}=\vec{r}-\vec{r}\,'$, $t(b,z)=\int_{z}^{\infty}
dz'\,n_{A}(b,z')$ and $n_{A}(\vec{r})$ is the matter density.
Furthermore, $\alpha_{pN}$ denotes the ratio of the real and imaginary
parts of the forward  $pN$ scattering amplitude and
$
\xi(\vec{\Delta}) =
\int d^2\vec{q} \;\
[d\sigma_{el}(pN)/d^2\vec{q}] \,
\exp(i \vec{q}\vec{\Delta}) \, .$
Since $k_{F}R_{A} \gg 1$, Eq.~(\ref{eq:4}) can be further simplified
by employing the local-density approximation [13],
$\rho_{1}(\vec{r},\vec{r}\,')=  \rho(\vec{\Delta})n_{A}(\vec{R}) \,$,
where $\vec{R}=(\vec{r}+\vec{r}\,')/2$.
The FSI has two effects: (1) besides the attenuation, there
appears a phase factor
$\exp[i\sigma_{tot}(pN)\alpha_{pN}n_{A}(b,z)(z-z')]$
which makes  the measured missing momentum $p_{m,z}$ different from
the longitudinal momentum $k_{z}$ of the target proton by an amount
\beq
k_{z}- p_{m,z}=  \Delta p_{m,z}
\approx
\sigma_{tot}(pN)\alpha_{pN}n_{A} \sim
\alpha_{pN}\cdot 70\, {\rm (MeV/c)} \, .
\label{eq:5}
\endeq
This momentum shift is the main contributor to distortion of the
$p_{m,z}$-distribution as compared to the PWIA where $p_{m,z}=k_{z}$.
All the other corrections are small [12].
(2) the factor $\exp\left[ t(\vec{b},z) \xi(\vec{\Delta})
\right]$ in the integrand of Eq.~(\ref{eq:4}) causes a broadening
of the $p_{\perp}$-distribution [6,8]:
\beq
f{_A}(\vec{p}_{\perp}) =
{1\over Z}\int d
E_{m}\, dp_{m,z}\, S(E_{m},p_{m,z},\vec{p}_{\perp}) =
\sum_{\nu = 0}^{\infty}
W^{(\nu)} f^{(\nu)}(\vec{p}_{\perp}) \, .
\label{eq:6}
\endeq
The probability, $W^{(\nu)}$, for $\nu$-fold elastic
rescattering equals
\beq
W^{(\nu)} = \frac{1}{A} \int dz d^2\vec{b} \;\ n_A(\vec{b},z)
\exp \left[-\sigma_{tot}(pN) t(\vec{b},z) \right]
{[t(\vec{b},z)
\sigma_{el}(pN)]^{\nu}\over \nu! }
\label{eq:7}
\endeq
and
\beq
f^{(\nu)}(\vec{p}_{\perp})=
\int d^2 \vec{s}\,[B/\nu\pi]
\exp \left( - B s^2/\nu \right)
f_{PWIA}(\vec{p}_{\perp} - \vec{s})\,\, ,
\eeq
where $f_{PWIA}(\vec{p}_{\perp})=\int dk_{z}n_{F}(k_{z},\vec{p}_{\perp})$
is the $p_{\perp}$-distribution in PWIA. For the purpose of the present
analysis it is sufficient to assume $f^{(0)}(\vec{p}_{\perp})=
f_{PWIA}(\vec{p}_{\perp})$ [12].
The normalization of $f_A$ is such that $\int d^{2}\vec{p}_{\perp}
f_{A}(\vec{p}_{\perp})= \sum_{\nu=0} W^{(\nu)}=T_{A}$, where
\arr
T_{A}=
{1\over A} \int dz d^2\vec{b} \;\ n_A(\vec{b},z)
\exp \left[-\sigma_{in}(pN) t(\vec{b},z) \right]\nonumber\\
=
{1\over A\sigma_{in}(pN)}\int d^{2}\vec{b}\,
\left\{1-\exp[-\sigma_{in}(pN)T(b)]\right\}
\label{eq:8}
\endarr
is total transmission factor or 'nuclear transparency factor' [2].
The quasielastic knock-out in parallel kinematics, $p_{\perp}=0$, is
dominated by the $\nu=0$ (PWIA) component of $f_{A}(\vec{p}_{\perp})$
and the corresponding transmission factor $W^{(0)}$ is given by
Eq.~(\ref{eq:8}) with $\sigma_{in}(pN)$ replaced by $\sigma_{tot}(pN)$.
The difference between $\sigma_{tot}(pN)$ and $\sigma_{in}(pN)$,
and consequently between $W^{(0)}$ and $T_{A}$, is very large at moderate
energies [14] (Fig.~1).

The above discussion proves that the distortion effects do not allow a
factorization of the measured spectral function
$S(E_{m},\vec{p}_{m})$
into $S_{PWIA}(E_{m},\vec{p}_{m})$ and an overall
attenuation factor. Experimentally, one compares the observed
nuclear cross section to the PWIA cross section integrated over
the acceptance domain $D$ in the $(E_{m},p_{m,z},p_{\perp})$ space [1].
In order to extract the attenuation effect, it
is therefore necessary to
include the shift $\Delta p_{m,z}$ in the PWIA
cross section, such that the
proper definition of nuclear transparency,
relevant to the NE18 experiment,
should be
\beq
T_{A}({\rm NE18})= {\int_{D} dE_{m}\, dp_{m,z}\,dp_{\perp}
S(E_{m},p_{m,z},p_{\perp}) \over  \int_{D} dE_{m}\, dp_{m,z}\,dp_{\perp}
S_{PWIA}(E_{m},p_{m,z}+\Delta p_{m,z},p_{\perp})}\,\, .
\label{eq:9}
\endeq
At values of $T_{kin}$ of the NE18 experiment the real part of the
$pN$ elastic amplitude is rather large, $\alpha_{pN}\sim -0.5$ [14],
and the effective shift (6) of the missing longitudinal momentum
$\Delta p_{m,z}  \sim -35 \,{\rm MeV/c}$. The effect of the
shift $\Delta p_{m,z}$ vanishes for a wide
$p_{m,z}$-acceptance, but for a narrow acceptance centered
at $p_{m,z}=p^{*}$ one has
\beq
{T_{A}(\Delta p_{m,z}) \over T_{A}(\Delta p_{m,z}=0)}
\approx 1 +
{5\over 2}\cdot {(\Delta p_{m,z}+p^{*})^{2}-(p^{*})^{2}
 \over k_{F}^{2} }
\label{eq:10}
\endeq
which enhances $T_{A}$ by $\sim 5\%$ at $p^{*}=0$. This is still
within the NE18 error bars.
At $Q^{2}\lsim 1\,{\rm (GeV/c)}^{2}$, {\sl i.e.,} at
$T_{kin}\lsim 0.5\, {\rm GeV}$, both $\sigma_{pN}$ and $\alpha_{pN}$
change rapidly with $T_{kin}$ [14] and the momentum shift may
lead to a spurious $Q^{2}$ dependence of the spectral function and
of the $y$-scaling function. This point has been  missed in the
discussion of FSI effects in [15].
The effects of the momentum shift (\ref{eq:5}) and
the $p_{\perp}$-broadening effect  approximately factorize.
The NE18 cut $p_{\perp} \lsim p_{max}=250\, {\rm MeV/c}$ [1]
partly includes struck protons which are elastically rescattered,
and we calculate $T_{A}(NE18)$ for this cutoff making use of
Eq.~(7) for the $p_{\perp}$-distribution.

As can be seen from Fig.~1 we find very good quantitative agreement
between our predictions
for $T_{A}({\rm NE18})$ and the data. This leads to
the following conclusions: (1) the Glauber
theory gives an adequate description of the FSI at moderately large
$Q^{2}$, (2) the signal of CT in $A(e,e'p)$ reactions
is still negligibly small
at $Q^{2} \leq 7$ (GeV/c)$^{2}$, in agreement
with our prediction ([2,3,5-7]). For $^{12}C(e,e'p)$ scattering
we estimate the threshold for CT at $Q^{2}\sim 5$ (GeV/c)$^{2}$
and CT effects shown as the short-dashed curve in Fig.~1
increase $T_{C}$ by $\sim 5\%$ at the largest $Q^{2}$ of
the NE18 experiment (Fig.~1). For heavier nuclei the CT signal is even
smaller. The main reason is the smallness of the contribution from
inelastic intermediate states because the amplitudes for
the diffraction excitations
$p+p\rightarrow i+p$ are small as compared to the amplitude
of elastic rescattering $p+p\rightarrow p+p$ [12,3,5,7].

In summary, we have calculated
the effects of distortions on the spectral
function for the nuclear transparency experiment NE18 at SLAC [1].
It has been shown that the FSI,
when treated in the Glauber approximation,
is able to explain the large shadowing effect observed in $d(e,e'p)$
reaction. Glauber's multiple-scattering theory is also found to
provide a good description of the
experimental results on heavier targets.
In spite of the small CT effect in
the data, even at the highest $Q^2$,
the possibility of observing a signal in dedicated $^{4}He(e,e'p)$
experiments at CEBAF should not be
ruled out. A discussion of this reaction
will be presented elsewhere [18].

\vspace{0.5cm}
\begin{center}
{\bf Acknowledgement}
\end{center}
Thanks are due to B.Fillipone, A.Lung and R.McKeown for discussions
and communications on the NE18 experiment and to H. Holtmann for
assistance in preparing the manuscript.
This work was supported in part by the Polish KBN grant 2 2409 9102
and the NSF grant PHY-89-21025.
\pagebreak


\pagebreak

{\bf \Large Figure captions:}

\begin{itemize}

\item[{\bf Fig.~1}]
   Predictions for the $Q^{2}$ dependence of the $p_{\perp}$-integrated
   nuclear transparency $T_{A}$ (dashed lines), the transparency
   $W^{(0)}$ for parallel kinematics (dotted lines) and the
   transparency $T_{A}({\rm NE18})$ including the acceptance cuts of the
   NE18 experiment [1]. The dot-dashed curve in panel for $^{12}C$
    indicates the
   effect from the onset of color transparency.

\end{itemize}
\end{document}